\documentclass[useAMS,usenatbib]{mn2e}
\pdfoutput=1
\usepackage{graphicx}

\def\aj{AJ}%
\def\actaa{Acta Astron.}%
\def\araa{ARA\&A}%
\def\apj{ApJ}%
\def\apjl{ApJ}%
\def\aap{A\&A}%
\def\aaps{A\&AS}%
\def\cjaa{Chinese J. Astron. Astrophys.}%
\def\mnras{MNRAS}%
\begin{document}

\title[The VMC survey - XVI.]{The VMC survey - XVI. Spatial
  variation of the cluster-formation activity in the innermost regions
  of the Large Magellanic Cloud\thanks{Based on observations obtained
    with VISTA at the Paranal Observatory under programme ID
    179.B-2003.}}

\author[A. E. Piatti et al.]{Andr\'es E. Piatti$^{1,2}$\thanks{E-mail: 
andres@oac.uncor.edu}, Richard de Grijs$^{3,4,5}$, Vincenzo Ripepi$^{6}$, Valentin D. Ivanov$^{7}$,
\newauthor Maria-Rosa L. Cioni$^{8,9,10}$, Marcella Marconi$^{6}$, Stefano Rubele$^{11}$,
\newauthor Kenji Bekki$^{12}$ and Bi-Qing For$^{12}$\\
$^1$Observatorio Astron\'omico, Universidad Nacional de C\'ordoba, Laprida 854, 5000, 
C\'ordoba, Argentina\\
$^2$Consejo Nacional de Investigaciones Cient\'{\i}ficas y T\'ecnicas, Av. Rivadavia 1917, 
C1033AAJ, Buenos Aires, Argentina \\
$^3$ Kavli Institute for Astronomy and Astrophysics, Peking University, Yi He Yuan Lu 5, Hai 
Dian District, Beijing 100871, China\\
$^4$ Department of Astronomy, Peking University, Yi He Yuan Lu 5, Hai Dian District, Beijing 
100871, China\\
$^5$ International Space Science Institute--Beijing, 1 Nanertiao,
Zhongguancun, Hai Dian District, Beijing 100190, China\\
$^6$ INAF, Osservatorio Astronomico di Capodimonte, via Moiariello 16, 80131 Napoli, Italy\\
$^7$ European Southern Observatory, Karl-Schwarzschild-Strasse 2, D-85748 Garching bei Munchen, Germany\\
$^8$ Universit\"at Potsdam, Institut f\"ur Physik und Astronomie, Karl-Liebknecht-Str. 24/25, 14476 Potsdam, Germany\\
$^9$ Leibnitz-Institut f\"ur Astrophysik Potsdam, An der Sternwarte 16, 14482 Potsdam, Germany\\
$^10$ University of Hertfordshire, Physics Astronomy and Mathematics, College Lane, Hatfield AL10 9AB, United Kingdom\\
$^{11}$ INAF, Osservatorio Astronomico di Padova, vicolo dell’Osservatorio 5, I-35122 Padova, Italy\\
$^{12}$ ICRAR, University of Western Australia, 35 Stirling Hwy, Crawley, WA 6009, Australia\\
}

\maketitle

\begin{abstract} 
We present results based on $YJK_{\rm s}$ photometry of star clusters
in the Large Magellanic Cloud (LMC), distributed throughout the
central part of the galaxy's bar and the 30 Doradus region. We
analysed the field-star decontaminated colour--magnitude diagrams of
313 clusters to estimate their reddening values and ages. The clusters
are affected by a mean reddening of $E(B-V) \in [0.2,0.3]$ mag, where
the average internal LMC reddening amounts to $\sim$ 0.1--0.2 mag. The
region covering 30 Doradus includes clusters with reddening values in
excess of $E(B-V)$ = 0.4 mag. Our cluster sample spans the age range
$7.0 \le \log(t$ yr$^{-1}) < 9.0$, represents an increase of 30 per
cent in terms of the number of clusters with robust age estimates and
comprises a statistically complete sample in the LMC regions of
interest here. The resulting cluster frequencies suggest that the
outermost regions of the LMC bar first experienced enhanced cluster
formation -- $\log(t$ yr$^{-1}) \in [8.5,9.0]$ -- before the activity
proceeded, although in a patchy manner, to the innermost regions, for
$\log(t$ yr$^{-1}) < 7.7$. Cluster frequencies in the 30 Doradus
region show that the area is dominated by very recent cluster
formation. The derived star-formation frequencies suggest that the
cluster and field-star populations do not seem to have fully evolved
as fully coupled systems during the last $\sim$ 100 Myr.
\end{abstract}


\section{Introduction}

The galaxies in the Local Group are uniquely suitable to verify the
results of analyses based on integrated cluster properties using
resolved stellar photometry
\citep[e.g.][]{dga06,colucci2012,baetal13,cezario2013,dmetal13}.
Dedicated photometric surveys of the Magellanic Clouds at both optical
and near-infrared wavelengths are now probing sufficiently deeply so
that we have a reasonable chance at resolving individual stars to well
below the main-sequence turn-off magnitudes characteristic of old
stellar populations. With our unparallelled access to deep
near-infrared observations obtained with the 4 m Visible and Infrared
Survey Telescope for Astronomy \citep[VISTA;][]{eetal06,es10} as part
of the VISTA near-infrared $Y, J, K_{\rm s}$ survey of the Magellanic
System (VMC) the star cluster systems in the Small and Large
Magellanic Clouds (SMC, LMC) can now be studied in detail to provide
unique insights into the properties of their resolved star cluster
populations.

In this paper, we address a long-standing issue of contention in
Magellanic Clouds studies, i.e., that of the coupling (if any) between
field-star and (massive) star cluster formation  (i.e., for
  cluster masses $M_{\rm cl} \ga 10^4$ M$_\odot$). The LMC exhibits a
well-known gap in the (massive) cluster age distribution between
$\sim$ 3 and 13 Gyr, while the age distribution of the field stellar
population appears more continuous. Numerous authors have asserted
that the LMC's field-star and star cluster formation histories are
significantly different \citep[e.g.][and references
  therein]{oetal96,getal98,s98}. The situation for the SMC is less
straightforward, although Rafelski \& Zaritsky (2005) provide
tentative evidence that the cluster and field-star age distributions
may also be significantly different in this system \citep[see
  also][]{gieles2007}. Similarly, the observed disparities between the
cluster and field-star age distributions in the (dwarf) starburst
galaxies NGC\,1569 \citep{aetal04} and M82 \citep{barker2008} seem to
offer evidence in support of a decoupling between star cluster and
field-star formation. If confirmed, this would be consistent with the
view that massive star cluster formation requires special conditions,
e.g., large scale gas flows, in addition to the presence of dense gas
\citep[cf.][]{az92,ee97,dgetal01}. However, this scenario may well
only apply to the most massive star clusters, whereas lower-mass
clusters could indeed be following the field-star formation history
more closely. This is our main scientific driver here.

 This is indeed an issue of general interest in the field of star
  and star cluster formation, given the commonly held notion that
  70--90\% of stars with masses in excess of 0.5 M$_\odot$ may form in
  clustered environments \citep[e.g.][]{ll03}. Since most
  present-day stars in galaxies like the Milky Way and the Magellanic
  Clouds form part of their host galaxies' field stellar population,
  it is not a stretch to suggest that a significant fraction of those
  field stellar populations originate from disrupted clusters, whether
  or not the latter were in fact gravitationally bound to begin with.

There is now significant evidence that star cluster systems appear to
be affected by a disruption mechanism that acts on very short
timescales ($\la 10$--30 Myr) and which may be mass-independent, at
least for masses $\ga 10^4$ M$_\odot$ \citep[e.g.][]{falletal05,bastianetal05,fall06}. 
This fast disruption mechanism, which is
thought to disrupt up to 50--90\% of the youngest clusters in a given
cluster population 
\citep[e.g.][]{ll91,whitmore04,bastianetal05,mengeletal05,gb06,whitmoreetal07}, 
is in essence caused by the rapid removal of the
intracluster gas on timescales of $\la 30$ Myr, a process coined
cluster `infant mortality' \citep{ll03}; it was originally
reported in the context of the number of very young embedded clusters
in the Milky Way, compared to their older, largely gas-free
counterparts.

Those clusters that survive the infant mortality phase will be subject
to the processes driving longer-term star cluster dissolution 
\citep[see for reviews][]{dgp07,degrijs10}. The longer-term
dynamical evolution of star clusters is determined by a combination of
internal and external timescales. These processes depend on the actual
cluster masses, with more massive clusters being more resilient to
disruptive influences. Based on these arguments, one would expect that
the lower-mass clusters in a given cluster population are the dominant
donors of stars to their host galaxies' field-star populations, while
the higher-mass cluster formation and evolution history is not
directly coupled to the field-star properties. In the context of the
LMC's star cluster population, cluster disruption has been the subject
of much debate. Nevertheless, and despite having access to cluster
samples exceeding 900 members, the jury is still out as regards the
typical time-scale of cluster disruption \citep[e.g.][]{pdg08} and whether 
this may vary as a function of position in the
galaxy \citep[e.g.][and references therein]{betal13,dgetal13}.

This paper is organised as follows. Section 2 provides an overview of
the data on which this study is based and the data reduction
procedures applied. We define our statistically complete LMC cluster
sample in Section 3 and use isochrone fits to determine the clusters'
best-fitting physical parameters in Section 4. Section 5 addresses the
key questions posed in this study; in particular, we consider the
cluster frequencies (CFs) across the LMC and compare the cluster
formation rates with the star-formation rates (SFRs) in the
corresponding field regions. We present our main conclusions in
Section 6.

\section{Data handling}

The VMC and its initial data are thoroughly described in
\citet{cetal11}, to which we refer the reader for details. Here we
used data of two VMC tiles located in the central part of the LMC bar
(LMC\,6\_4) and in a region covering the 30 Doradus star-forming
region (LMC\,6\_6), for which the VISTA imaging campaign has been
completed. These regions are of significant scientific interest, 
since they contain a large number of star clusters which can provide 
statistical results about the central region of the LMC
where other tiles are still being observed. Indeed, the purpose of this
paper is to present a photometric analysis of the catalogued clusters
located in those regions based on the most complete VMC data
set to date. The available photometric data allow us to confirm the
physical reality of the catalogued star clusters and estimate their 
fundamental parameters. We also investigate
the CF in the regions of interest to assess whether star
clusters and field stars have evolved as a coupled system.
Figure 1 shows the location of all LMC tiles of the VMC
survey. The red rectangles mark those used in this paper.

\begin{figure}
\includegraphics[width=84mm]{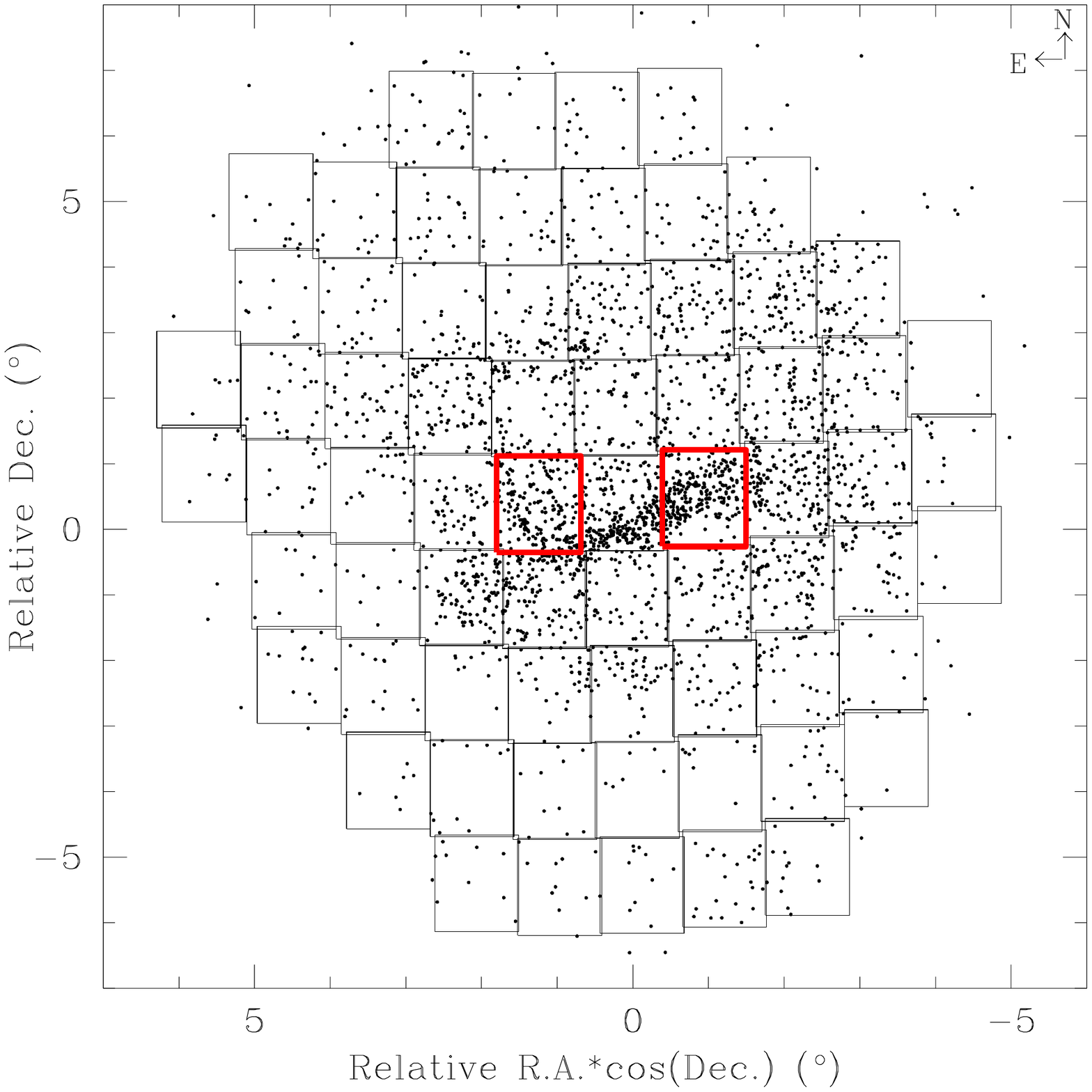}
\caption{Spatial distribution of the \citet{betal08}'s catalogue of
  star clusters in the LMC centred at R.A. = 05$^{\rm h}$ 23$^{\rm m}$
  34$^{\rm s}$, Dec. = $-$69$\degr$ 45$\arcmin$ 22$\arcsec$ (J2000),
  projected onto the sky. The objects studied in this paper are
  located in tiles highlighted as red rectangles: LMC\,6\_4 (right)
  and LMC\,6\_6 (left). The remaining VMC tiles across the LMC are shown
as black rectangles.}
\label{fig1}
\end{figure}

The tile LMC\,6\_4 and 6\_6 data refers to observations acquired from
October 2010 until November 2011 and from November 2009 to October
2010, respectively. Seeing constraints, imposed for the purpose of
homogenizing crowded and uncrowded field observations, range between
1.0 and 1.3 arcsec ($Y$), 0.9 and 1.2 arcsec ($J$), and 0.8 and 1.0
arcsec ($K_s$) and may exceed those values by 10 per cent according to
observing policy.  One tile covers uniformly an area of $\sim$1.5
deg$^2$, as a result of the mosaic of six paw-print images, in a given
waveband with 3 epochs at $Y$ and $J$, and $12$ epochs at
$K_s$. Individual epochs have exposure times of 800 s ($Y$ and $J$)
and 750 s ($K_s$). We used the v1.3 VMC release paw-prints.  They were
processed by the VISTA Data Flow System's \citep[VDFS;][]{eetal04}
pipeline \citep{ietal04} and retrieved from the VISTA Science Archive
\citep[VSA;][]{hetal04}. The processed paw-print images were used to
derive the effective point spread functions (PSFs) using the {\sc
  iraf/daophot} routines \citep{setal90}. We generated a reference
PSF, which was convolved with the paw-print images to homogenize the
resulting PSFs. We repeated these steps for each epoch
separately. Finally, all homogenized paw-print images were combined
using the SWARP tool \citep{betal02}, as described in \citet{retal12},
thus generating deep tile images with homogeneous PSFs.

We performed PSF photometry using the {\sc iraf/daophot} package
to generate photometric catalogues which return right ascension and
declination, object magnitude, its error and sharpness, local
magnitude completeness and number of stars used to compute the latter
in the $Y,J,K_{\rm s}$ passbands for $\sim$2.6 million of sources in
each tile.
% LMC\,6\_4 and
%6\_6, respectively. 
We used the {\sc psf} task to produce the PSF model (which varies
across the sky) and the {\sc allstar} task to perform our photometry,
using a radius of three pixels. We checked that our PSF photometry
produced results consistent with those provided by the VSA for the
bulk of the observed stars \citep{retal12,retal15}. The final
photometric catalogues are publicly available both at the
VSA\footnote{http://horus.roe.ac.uk/vsa/} and
ESO\footnote{http://www.eso.org/sci/observing/phase3/data\_releases.html}
science archives.

We ran a large number of artificial-star tests to estimate the
incompleteness and error distribution of our photometric catalogues
for each tile and in every part of the colour--magnitude diagram
(CMD), as described in \citet{retal12}, where the photometry for two
small subregions in tile LMC\,6\_6 is presented. Photometric errors of
0.10 mag were derived for stars with $Y$ = 20.8 mag, $J$ = 20.6 mag
and $K_{\rm s}$ = 19.7 mag in tile LMC\,6\_4 and for $Y$ = 21.2 mag,
$J$ = 20.9 mag, and $K_{\rm s}$ = 19.9 mag in tile LMC\,6\_6. The 50
per cent completeness level is reached at $Y$ = 19.9 mag, $J$ = 19.5
mag and $K_{\rm s}$ = 19.6 mag in tile LMC\,6\_4 and at $Y$ = 20.4
mag, $J$ = 20.1 mag, and $K_{\rm s}$ = 19.9 mag in tile LMC\,6\_6,
throughout the entire tile areas.  Photometric completeness is used
here as a reference to our knowledge of the faintest cluster
main-sequence turnoff (MSTO) reachable by our photometry.

\section{The cluster sample}

Recognizing catalogued star clusters in deep VMC tile images is
neither straightforward nor simple. The catalogued
objects were originally identified from optical images (e.g., from
Digitized Sky Survey images; DSS\footnote{The Digitized Sky Surveys
  were produced at the Space Telescope Science Institute under
  U.S. Government grant NAG W-2166. The images of these surveys are
  based on photographic data obtained using the Oschin Schmidt
  Telescope on Palomar Mountain and the UK Schmidt Telescope. The
  plates were processed into the present, compressed digital form with
  the permission of these institutions.}) which sometimes look rather
different compared with their appearance at near-infrared
wavelengths (bright red stars may dominate the cluster light while faint 
blue stars may be entirely absent in the near-infrared). 
In addition, the spatial resolution and depth of the
images on which the clusters were identified differ from the
equivalent parameters pertaining to the VMC images. Thus, for
instance, single relatively bright stars might look like an unresolved
compact cluster in images of lower spatial resolution, or unresolved
background galaxies could be mistaken for small star clusters in
shallower images. Offsets in the compiled coordinates with respect to
the objects’ centres cannot be ruled out either.

To avoid mismatches between observed objects and the actual list of
catalogued clusters, we first overplotted the positions of the
catalogued clusters on the deepest stacked $K_{\rm s}$ image. This
way, based on using the coordinates resolved by the
SIMBAD\footnote{http://simbad.u-strasbg.fr/simbad/} astronomical data
base, we unambiguously recognized the observed clusters one by one in the
$K_{\rm s}$ image. Note that the main aim of this task is to confirm
the compiled cluster coordinates and sizes so as to extract from the
VMC photometric catalogues the magnitudes of the stars in the cluster
region. We are not interested in properties such as the clusters'
structures, stellar density profiles or radii, but we aim at obtaining
accurate photometry for those stars which allow us to meaningfully
define the clusters' fiducial sequences in their CMDs.

In order to build a cluster catalogue that is as complete as possible,
we made use of those previously compiled by \citet[][hereafter
  PU00]{pu00}, \citet[][hereafer B08]{betal08} and \citet[][hereafter
  G10]{getal10}. Particularly, B08 provided angular dimensions, which
we used for assessing whether such regions contain sufficient numbers
of stars to construct reliable cluster CMDs with the smallest fraction
of unavoidable field-star contamination. While recognizing the
catalogued clusters on the $K_{\rm s}$ tile image, we discarded those
in the line-of-sight of H{\sc ii} regions, unresolved clusters, those
for which the VMC photometry does not reach the MSTO or where the
observed cluster CMDs do not show any cluster sequences. While
attempting to cross-correlate the cluster names and their coordinates,
we could not identify BSDL\,1161 in the $K_{\rm s}$ tile image, and we
found that SL\,368 is listed at incorrect coordinates in B08 (and,
hence, in SIMBAD as well). Nevertheless, G10 compiled the correct
cluster loci. Our final cluster sample includes 313 catalogued
clusters.

\section{Star cluster parameters}

In general terms, the observed CMDs of the selected objects are the
result of the superposition of different stellar populations
distributed along the line of sight. For this reason, using the
observed CMDs without subtracting the luminosity function and colour
distribution of field stars may lead to incorrect
interpretations. Moreover, since the catalogued clusters were
initially identified as small concentrations of stars on the basis of
stellar density fluctuations, their real physical nature requires
subsequent confirmation. In order to disentangle cluster stars from
field stars, we used CMDs of adjacent fields to subtract the local LMC
field luminosity function and colour distribution.

We constructed cluster CMDs based on all measured stars distributed
within a circle with a radius of three times that tabulated by B08
based on visual inspection of the objects on DSS images or by us from
the deepest $K_s$ images. The objects are of small angular size,
typically $\sim$ 0.6 arcmin ($\sim$ 8.7 pc) in diameter.  These
regions are sufficiently large to encompass most of the cluster
regions. Note that our main aim here is to clean the cluster CMDs from
contamination by field stars within an area around the clusters'
centres that is nine times larger than $\pi$$r$$_{\rm cluster}^2$, so
that we do not need to trace their radial profiles. Once the cluster
areas were delineated, we defined four additional regions with areas
equal to the cluster regions and located more or less equidistant from
the clusters' centres. For each of the latter regions we constructed
CMDs showing the defining features of the local LMC star field in that
particular direction.

We next applied a procedure which was designed to compare each of the
field CMDs to the cluster CMD and subtracted from the latter a
representative field CMD in terms of stellar density, luminosity
function and colour distribution. This was done by comparing the
numbers of stars counted in boxes distributed in a similar manner
throughout all CMDs. The boxes were allowed to vary in size and
position throughout the CMDs in order to meaningfully represent the
actual distribution of field stars. For additional details, see
\citet{petal14b,petal15a}, who applied this same procedure to other
VMC data sets. Figure 2 illustrates the performance of the cleaning
procedure for NGC\,2108, where we plotted three different CMDs: that
for the stars located within the cluster radius (top left-hand panel),
a single-field CMD for an annulus -- outer and inner radii equal to
3.163 and 3.0 times the cluster radius -- centred on the cluster, as
well as the cleaned cluster CMD (bottom left-hand panel). The
schematic diagram with a superimposed circle of radius equal to the
cluster radius is shown in the bottom right-hand panel. The pink,
light and dark blue filled circles in the bottom panels represent
stars with cluster membership probabilities $P$ $\le$ 25 per cent, $P$
= 50 per cent and $P$ $\ge$ 75 per cent, respectively. The 313
individual photometric catalogues for the studied clusters are
provided in the online version of the journal. The columns of each
catalogue successively lists the magnitude, the error and the
sharpness for every source within a circle of radius three times that
of the cluster and centred on the cluster's centre, in $Y$, $J$ and
$K_s$, respectively, the R.A. and Dec. and the photometric membership
probability ($P$). The latter is encoded with numbers 1, 2, 3 and 4 to
represent probabilities of 25, 50, 75 and 100 per cent, respectively.

\begin{figure}
\includegraphics[width=84mm]{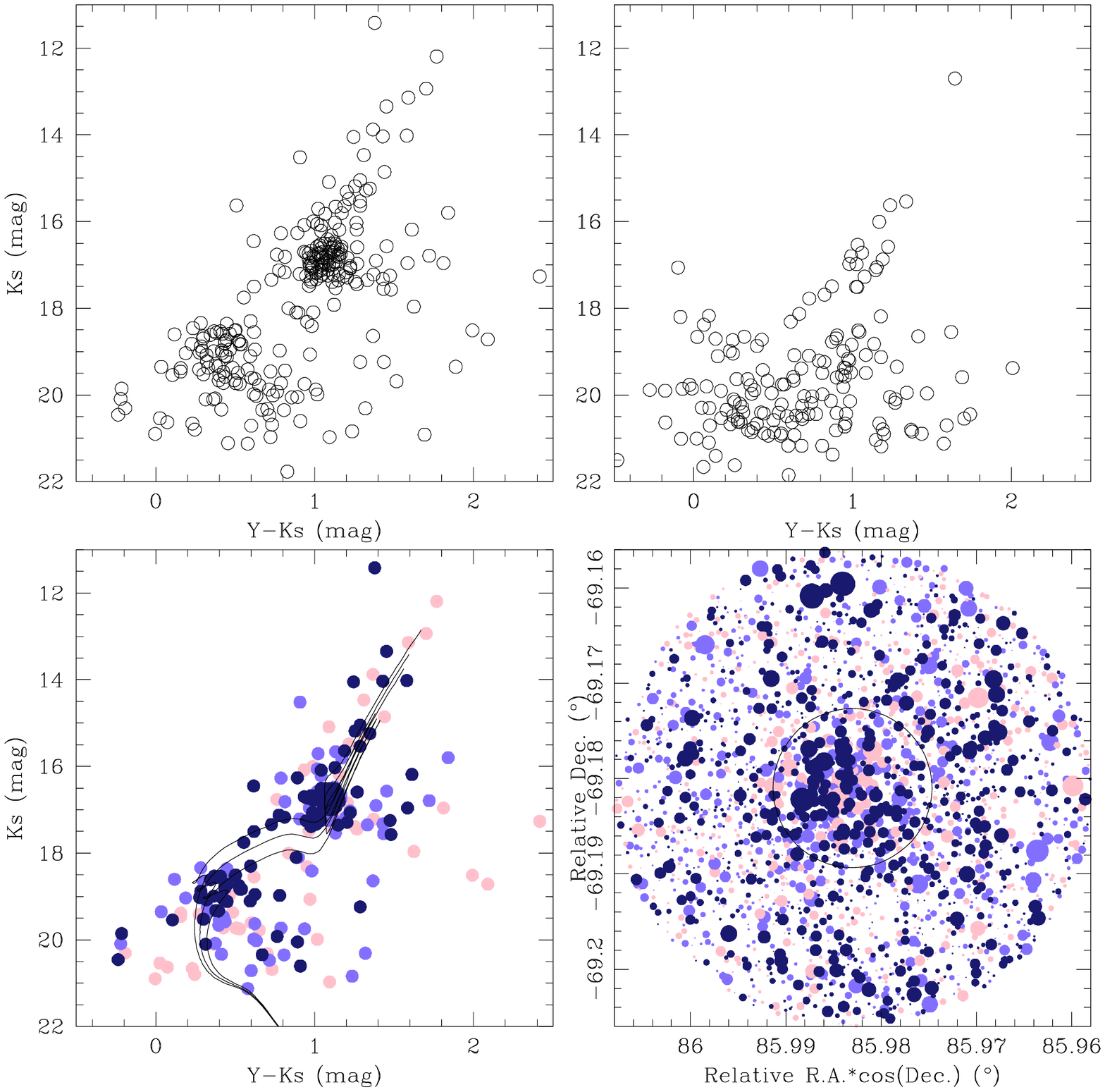}
\caption{CMDs for stars in the field of NGC\,2108: the observed CMD
  composed of the stars distributed within the cluster radius (top
  left-hand panel); a field CMD for an annulus -- outer and inner
  radii equal to 3.163 and 3.0 times the cluster radius -- centred on
  the cluster (top right-hand panel); the cleaned cluster CMD (bottom
  left). Colour-scaled symbols represent stars that statistically
  belong to the field ($P \le$ 25\%, pink), stars that might belong to
  either the field or the cluster ($P =$ 50\%, light blue), and stars
  that predominantly populate the cluster region ($P \ge$ 75\%, dark
  blue). Three isochrones from \citet{betal12} for log($t$ yr$^{-1}$)
  = 8.8, 8.9, and 9.0 and $Z$ = 0.006 are also superimposed.  The
    lack of fainter main-sequence stars results in less satisfactory
    matching at fainter magnitudes. The schematic diagram centred on
  the cluster for a circle of radius three times the cluster radius is
  shown in the bottom right-hand panel. The black circle represents
  the adopted cluster radius. Symbols are as in the bottom left-hand
  panel, with sizes proportional to the stellar brightnesses. North is
  up; East is to the left.}
\label{fig2}
\end{figure}

We estimated reddening values and ages for our 313 sample clusters
using the theoretical isochrones of \cite{betal12} in the Vegamag
system (where, by definition, Vega has a magnitude of zero in all
filters), corrected by $-0.074$ mag in $Y$ and $-0.003$ mag in $K_{\rm
  s}$ to put them on the VMC system \citep{retal15}, to match the
cleaned cluster CMDs. We adopted the same distance modulus for all
clusters $(m-M)_0 = 18.49 \pm 0.09$ mag \citep{dgetal14} and $K_{\rm
  s}-M_{K_{\rm s}} = (m-M)_0 + 0.372 E(B-V)$, for $R_V = 3.1$
\citep{cetal89,getal13}, since by considering an average depth for the
LMC disc of (3.44$\pm$1.16) kpc \citep{ss09}, we derived a smaller age
difference than that resulting from the isochrones (characterized by
the same metallicity) bracketing the observed cluster features in the
CMD.

We used isochrones for $Z= 0.006$ ([Fe/H] $= -0.4$ dex), which
corresponds to the mean LMC cluster metal content for the last $\sim$
2--3 Gyr \citep{pg13}. Additionally, the $Y-K_{\rm s}$ colour is not
sensitive to metallicity differences smaller than $\Delta$[Fe/H] $\sim
0.4$ dex, which is adequate given the spread of the stars in the CMDs
\citep[see][]{petal15a}. Indeed, we tried using isochrones with
metallicities [Fe/H] = 0.0 dex and [Fe/H] = $-0.7$ dex and found
negligible differences with respect to that of [Fe/H] $= -0.4$ dex,
keeping in mind the relatively sparse nature of the majority of our
clusters and the intrinsic spread of the stars in the CMDs. We found
that isochrones bracketing the derived mean age by $\Delta E(B-V) =
\pm 0.05$ mag and $\Delta \log(t$ yr$^{-1}) = \pm 0.1$ represent the
overall uncertainties associated with the observed dispersion in the
cluster CMDs. Although the dispersion is smaller in some cases, we
prefer to retain the former values as an upper limit to our error
budget. Table 1 lists the diameters used and the resulting reddening
and age values obtained for the entire cluster sample, while the
bottom left-hand panel of Fig. 2 illustrates the results of the
isochrone matching for NGC\,2108.

\begin{table}
\caption{Fundamental parameters of the LMC clusters studied here.  
We assume a distance modulus of 18.49 mag and a metallicity of
[Fe/H]$= -0.4$ dex. The complete table is available online as Supporting
Information.}             
\begin{tabular}{@{}lccccc}\hline
Name & R.A. & Dec. & $d$ &$E(B-V)$ & log($t$ yr$^{-1}$) \\
     & (\degr) & (\degr) & (arcmin) & (mag) &   \\\hline
... & ... & ... & ... & ... & ... \\ 
NGC\,1858 &      77.500 &  -68.904 & 0.70 & 0.30 &  7.30\\
HS\,335   &      82.737 &  -69.341 & 0.40 & 0.10 &  8.60\\
NGC\,2009 &      82.750 &  -69.182 & 0.80 & 0.15 &  7.50\\
... & ... & ... & ... & ... & ...\\ \hline
\end{tabular}
\end{table}

The resulting reddening distributions for the clusters located in
tiles LMC\,6\_4 and 6\_6 are shown in Fig. 3, where we have also
included an inset showing their spatial distribution. We converted
$E(Y-K_{\rm s})$ to $E(B-V)$ using $E(Y-K_{\rm s}) = 0.84 E(B-V)$
\citep{cetal89}. Both  surveyed regions are characterized by a
mean reddening spanning the range $E(B-V) \in [0.2,0.3]$ mag. The tile
centred on the LMC bar contains a slightly larger number of clusters
with reddening values towards the upper limit of this range. On the
other hand, the tile covering 30 Doradus includes clusters with
reddening values in excess of $E(B-V)$ = 0.4 mag. Bearing in mind that
the observed cluster sequences are affected by interstellar extinction
which is the sum of Milky Way (foreground) and LMC (internal)
reddening, and that the mean Galactic reddening towards the LMC is
$E(B-V) \approx 0.08$ mag \citep{detal01,sf11}, it follows that the
average (internal) reddening in these particular LMC regions is $\sim$
0.1--0.2 mag. This range is in very good agreement with a number of
previous estimates \citep[e.g.,][]{oetal95,ss09,tetal13}, although
slightly lower values have also been found from analysis of field
red-clump and variable stars \citep{cetal03,hetal11}.
%The spatial
%reddening distribution reveals the bar region as more heavily reddened
%on average, and that the 30 Doradus region exhibits the highest
%$E(B-V)$ values towards certain specific star-forming regions.

\begin{figure*}
\includegraphics[width=144mm]{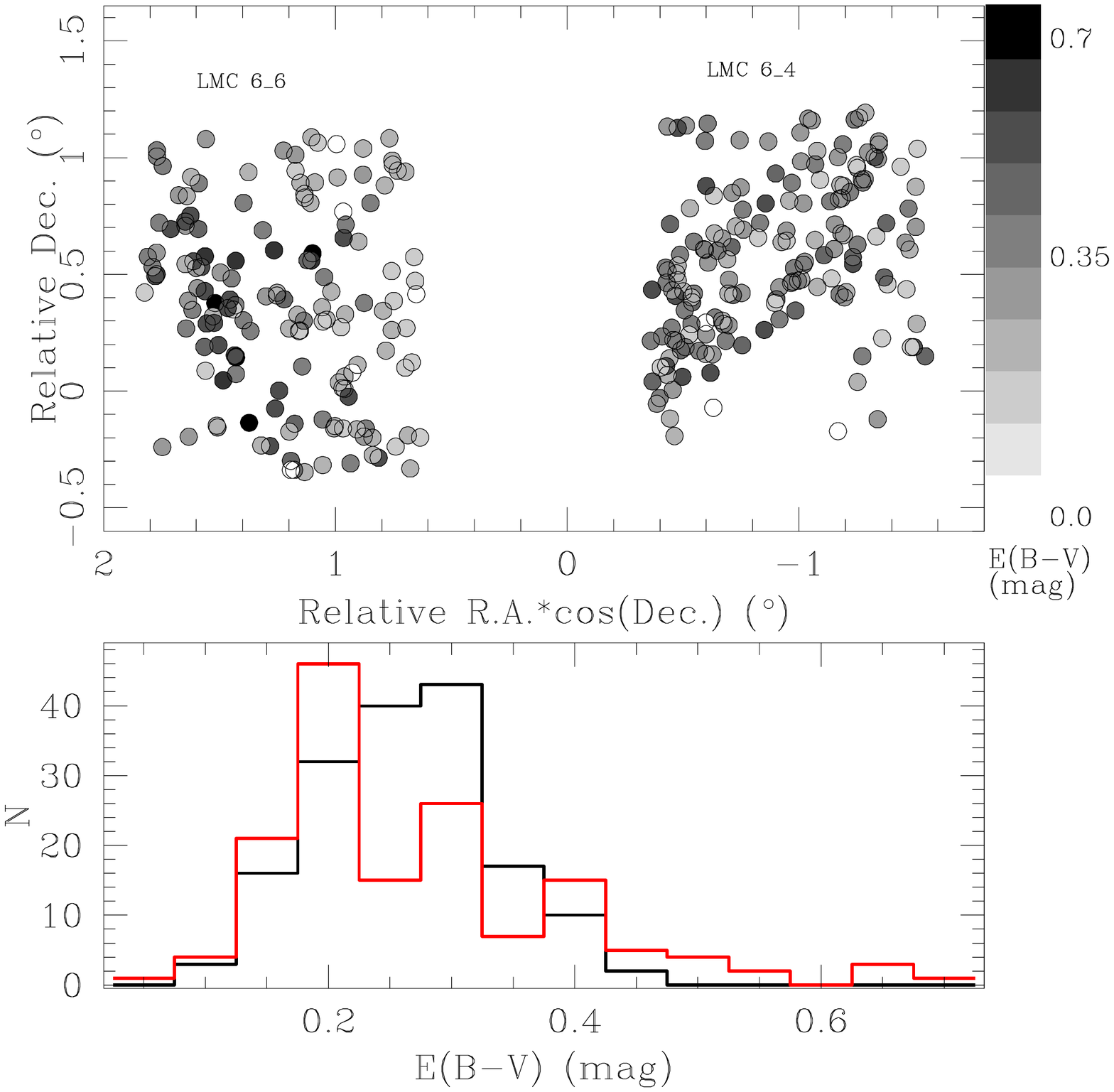}
\caption{Reddening histograms for tiles LMC\,6\_4 (black) and
  LMC\,6\_6 (red) (bottom panel) and the spatial distribution of the
  clusters' reddening, with darker shading representing higher
  reddening values (top panel). The grey scale is in units of $E(B-V)$
  (mag).
%Note that tiles are square although they appear as rectangles.
}
\label{fig3}
\end{figure*}

\begin{figure}
\includegraphics[width=84mm]{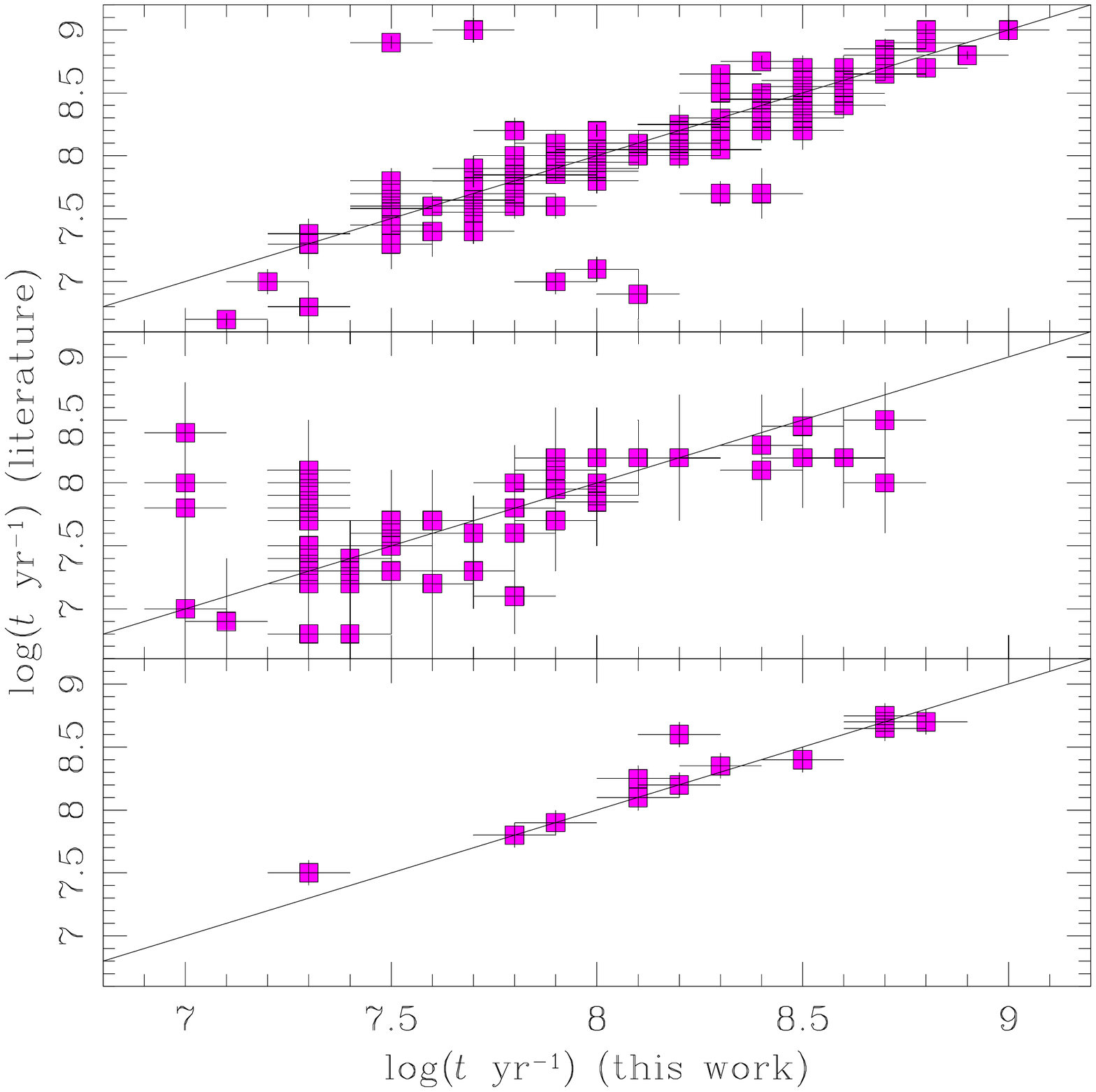}
\caption{Comparison of the cluster age estimates derived here with
  those from the literature: PU00 (top panel), G10 (middle panel) and
  Washington photometry \citep[][]{p12c,p14,chetal15} (bottom
  panel). Error bars and the locus of equality are also shown.}
\label{fig4}
\end{figure}

\begin{figure*}
\includegraphics[width=144mm]{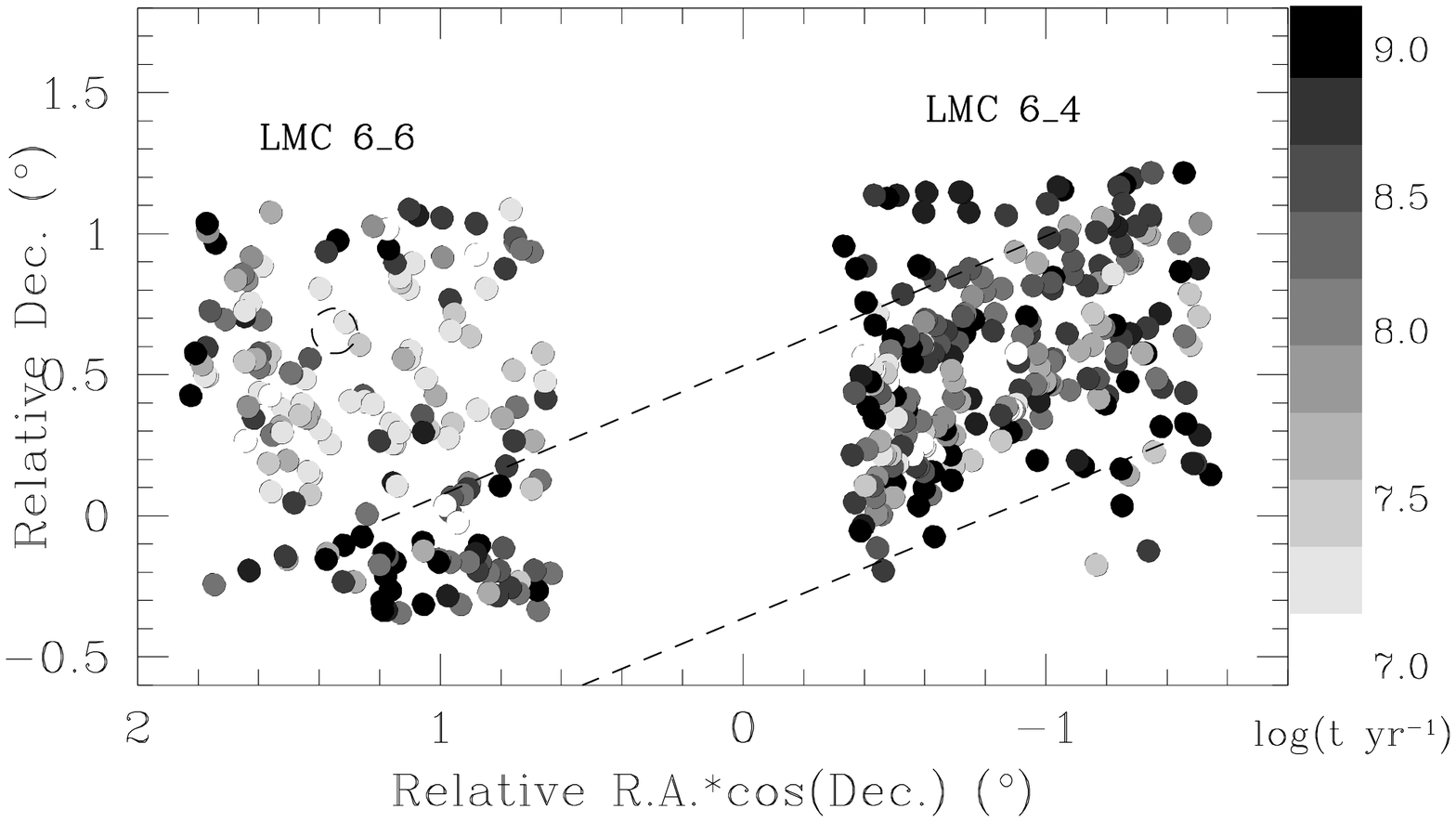}
\caption{Spatial distribution of the cluster ages, with darker shading
  representing older ages, centred at R.A. = 05$^{\rm h}$ 23$^{\rm m}$
  34$^{\rm s}$, Dec. = $-$69$\degr$ 45$\arcmin$ 22$\arcsec$ (J2000).
  North is up; East is to the left. The grey scale is in units of
  $\log(t$ yr$^{-1})$.  The locations of 30 Doradus and the LMC
    bar are indicated with a dashed circle and two parallel lines
    \citep{ss09b}, respectively.
%Note that tiles are square although they appear as rectangles.
}
\label{fig5}
\end{figure*}

Figure 4 shows a comparison of the cluster age estimates derived here
with those previously obtained from similar isochrone matching.  The
latter assumed a constant metal content for the LMC of [Fe/H] $= -0.4$
dex, a distance modulus of 18.50 mag (except PU00 who adopted 18.24
mag and mentioned that change in the distance modulus by about 0.15
mag produced difference in ages of less than a few percent, which is
much less than the derived errors), and isochrones computed by the
Padova group \citep{betal94,metal08}.  Agreement with the OGLE-based
age estimates  (200 clusters) is good, just like that with the
cluster ages determined on the basis of Washington photometry 
  \citep[][3, 7 and 3 clusters, respectively]{p12c,p14,chetal15}. The
most discrepant points in the OGLE age comparison are associated with
clusters whose CMDs include relatively bright blue stars that PU00
either rejected or included in the isochrone matching, thus resulting
in, respectively, older (HS\,223, $\log(t$ yr$^{-1}) =8.9$; BSDL\,725,
$\log(t$ yr$^{-1}) =9.0$) or younger (HS\,362, $\log(t$ yr$^{-1})
=6.9$; HS\,219, $\log(t$ yr$^{-1}) =7.0$; H\,88-169, $\log(t$
yr$^{-1}) =7.1$) ages than our values. As for the comparison with the
ages based on Washington photometry, which were obtained from
isochrone matching to CMDs previously corrected for field-star
contamination using the same technique described above and for
clusters not studied by either PU00 or G10, better agreement can still
be achieved. The correlation with the ages determined by G10 (89
  clusters) is less tight, because G10 did not perform field-star
decontamination, although these authors mention that field
contamination is a severe effect in the extracted cluster CMDs and
may, therefore, affect the age estimates significantly. Consequently,
their age estimates likely reflect the composite LMC stellar
populations (e.g., KMK\,88-74 ($\log(t$ yr$^{-1}) =7.8$), BSDL\,2807,
($\log(t$ yr$^{-1}) =8.0$), KMK\,88-86 ($\log(t$ yr$^{-1}) =8.4$)).
This possibility alerts us to the fact that solely the circular
extraction of the observed CMDs of clusters located in highly
populated stellar fields is not sufficient for accurate isochrone
matching of the main cluster features \citep[see
  also][]{pb12,p14,petal14b,petal15a}.

Based on Fig. 4, we conclude that the age scale derived in this paper
is consistent with those pertaining to our comparison samples. We can
therefore safely merge these samples to compile a larger cluster
sample characterized by an age distribution put on an homogeneous
scale. Of our 313 sample clusters, 95 did not have any previous age
estimates -- a $\sim$ 30 per cent increase of clusters with known ages
in these fields.
%Our derivation of new ages for this subsample represents an
%increase of $\sim$ 30 per cent of clusters with known ages in these
%fields. 
The final cluster age catalogue is composed of our 313 age estimates,
combined with 74 clusters from PU00, five clusters from G10 and 15
clusters from other studies (mainly based on Washington
photometry).  Whenever a cluster has more than one age estimate,
  we adopted a weighed average value. All ages considered here were
derived based on isochrone matching of the cluster CMDs, spanning the
age range $7.0 < (\log(t$ yr$^{-1}) < 9.0$. Figure 5 shows the
clusters' spatial distribution, with darker points representing older
ages. Note the obvious distribution of very young clusters around 30
Doradus in tile LMC\,6\_6, as well as of young clusters in the
innermost regions of the bar in tile LMC\,6\_4.

\section{The cluster frequencies in tiles LMC\,6\_4 and 6\_6}

A number of different studies have shown that the LMC's field SFR 
  is variable across the galaxy \citep[][hereafter
  HZ09]{retal12,meschin14,hz09}. In particular, HZ09 concluded from
the concordance between the star-formation and chemical-enrichment
histories of the field and cluster populations that the field and
cluster star-formation modes are tightly coupled. Recently,
\citet{pg13} studied 21 LMC regions spread across the main body of the
galaxy and found that the cluster and field star age--metallicity
relationships (AMRs) of such regions show a satisfactory match only
for the last 3 Gyr, i.e., $\log(t$ yr$^{-1}) \le9.5$, while for the
oldest ages -- $t>$11 Gyr, $\log(t$ yr$^{-1})>10.05$ -- the cluster
AMR represents a lower envelope to the field AMR.

We are interested in constructing the CFs, i.e., the number of
clusters formed per unit time interval as a function of age, as a
function of position in both tiles studied here. It would appear
reasonable to infer that if the star-formation history of LMC field
stars has been different throughout the galaxy, and field stars and
clusters show some evidence of common formation and
chemical-enrichment histories, then the CFs should reflect the same
spatial variations as seen in the field stars. \citet[][hereafter
  P14]{p14b} showed that the number of catalogued clusters without age
estimates is relatively small with respect to the total number of
catalogued clusters in the LMC. He also showed that nearly half of the
more massive clusters in his sample with age estimates mostly trace
the overall behaviour of the CFs. Alternatively, clusters with masses
below $10^3$ M$_{\odot}$ for ages younger than 1 Gyr do not contribute
significantly to the normalized CFs, although they represent on
average nearly half of the total number of clusters with age
estimates. In addition, he found that different model age
distributions for the total number of clusters without age estimates
do not affect the resulting CFs. Moreover, he concluded that possible
incompleteness of non-catalogued clusters (which include the faintest
clusters in the galaxy) does not play an important role in the
resulting CFs either. Since our cluster age catalogue for the two LMC
tiles of interest contains $\sim$ 30 per cent more clusters than
previously known, we are confident that we can consider our results
statistically robust. Indeed, our cluster sample includes $\sim$ 80
per cent of the catalogued clusters in the surveyed regions.

In constructing the CFs, we first divided the tiles into 12 subregions
of similar size, and then counted the number of clusters in each
region, taking into account bin size and age uncertainty effects.
This procedure allowed us to build age histograms (and hence CFs)
which best reproduce the intrinsic age distribution of each
subregion. For additional details about the treatment of bin size and
age uncertainty effects, see, e.g., \citet{pg13}, P14 and
\citet{petal15a}. All CFs were normalized to the total number of
clusters included in our sample. We assume that there has been no
statistically significant mixing of clusters for ages of $\log(t$
yr$^{-1}) < 9.0)$ (which represents less than the last 10 per cent of
the galaxy's lifetime) between regions. Indeed, \citet{vdmk14} showed
that for most tracers (including clusters and field stars), the LMC's
line-of-sight velocity dispersion is at least a factor of $\sim$2
smaller than the corresponding rotation velocity, which implies that
the LMC as a whole is a kinematically cold disc system.
%Additionally, the distances and metallicities
%associated with the clusters tell us that most likely they were formed
%in the inner LMC region. 
Figures 6 and 7 show the resulting CFs for the subregions of tiles
LMC\,6\_4 and 6\_6, respectively. We include insets with the schematic
distribution of the 12 subregions and a colour reference of the
respective CF curves.  From Fig. 6, it is clear that the outermost
subregions of the LMC bar, which crosses tile LMC\,6\_4 along the
SE--NW diagonal (blue lines), have experienced more prominent cluster
formation in the past -- $\log(t$ yr$^{-1}) \sim 8.5$--9.0 -- and more
recently, $\log(t$ yr$^{-1}) < 7.7$, cluster formation in the
innermost regions has become more active (red and green
lines). However, the cluster formation scenario in the central parts
of the bar appears to be rather complex and patchy. In tile LMC\,6\_6
(Fig. 7) the most striking role is played by the 30 Doradus region,
which dominates the recent cluster formation activity (dark green and
black lines). Likewise, 30 Doradus is spatially surrounded by cluster
formation activity characterized by older ages with increasing
distance from the object (blue and red lines), possibly approaching
the intrinsic CF in the outer LMC bar (see fig. 5 in P14).

\begin{figure*}
\includegraphics[width=144mm]{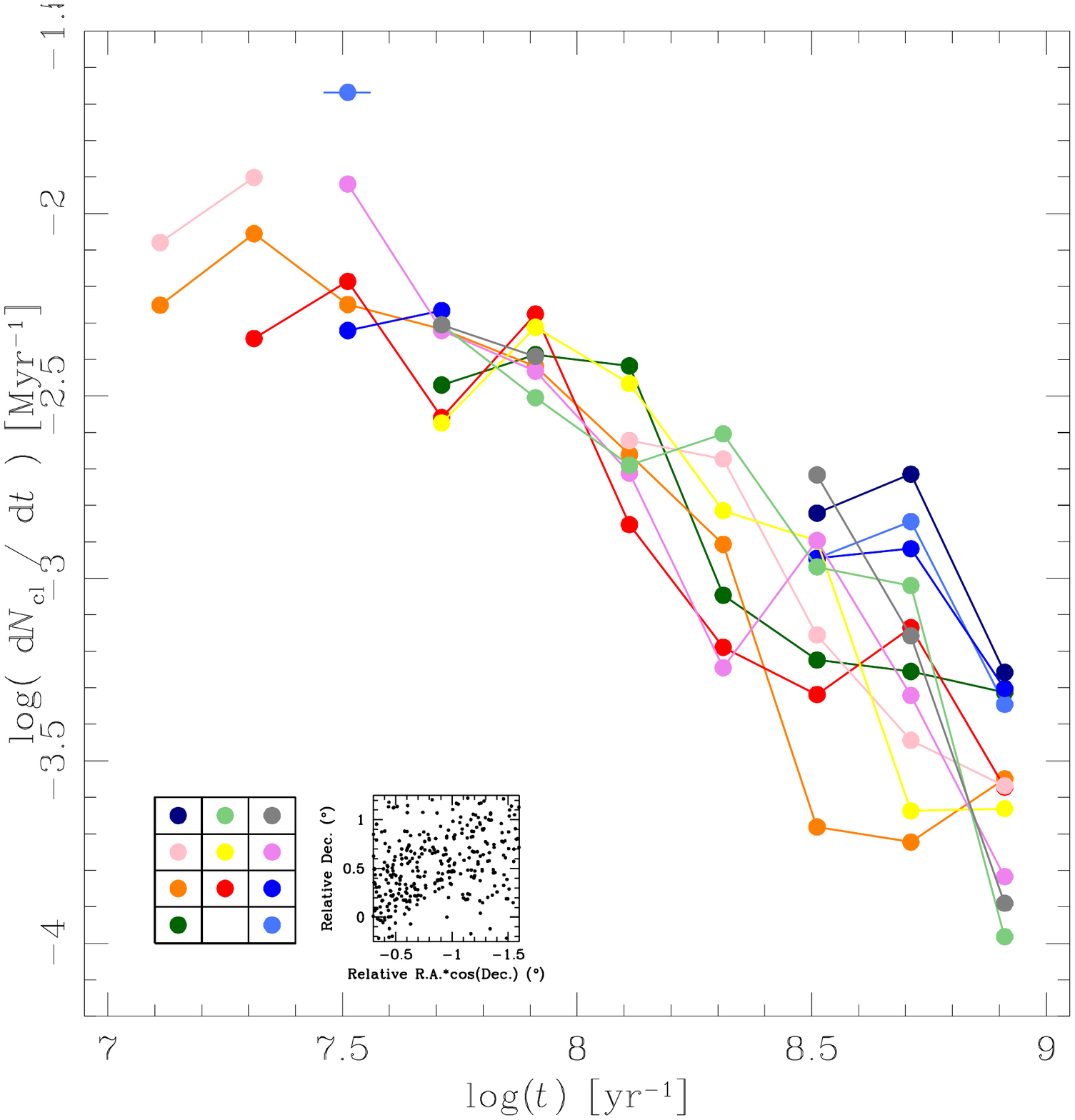}
\caption{CFs for subregions in tile LMC\,6\_4. The bottom 
    right-hand panel shows the spatial distribution of
    \citet{betal08}'s catalogue of star clusters, while the left-hand
    panel schematically represents the subregions and indicates the
  colour used in drawing the respective CFs. North is up and east is
  to the left.  Note that computed values (filled circles) for each CF
  are connected by a solid line. Some CFs do not have points for
  certain age intervals because of the lack of clusters within them,
  which resulted in truncated CFs. In such cases, we drew isolated
  points with a short line to represent that they belong to a
  truncated CF. One subregion does not contain any cluster in our
  sample.}
\label{fig6}
\end{figure*}

\begin{figure*}
\includegraphics[width=144mm]{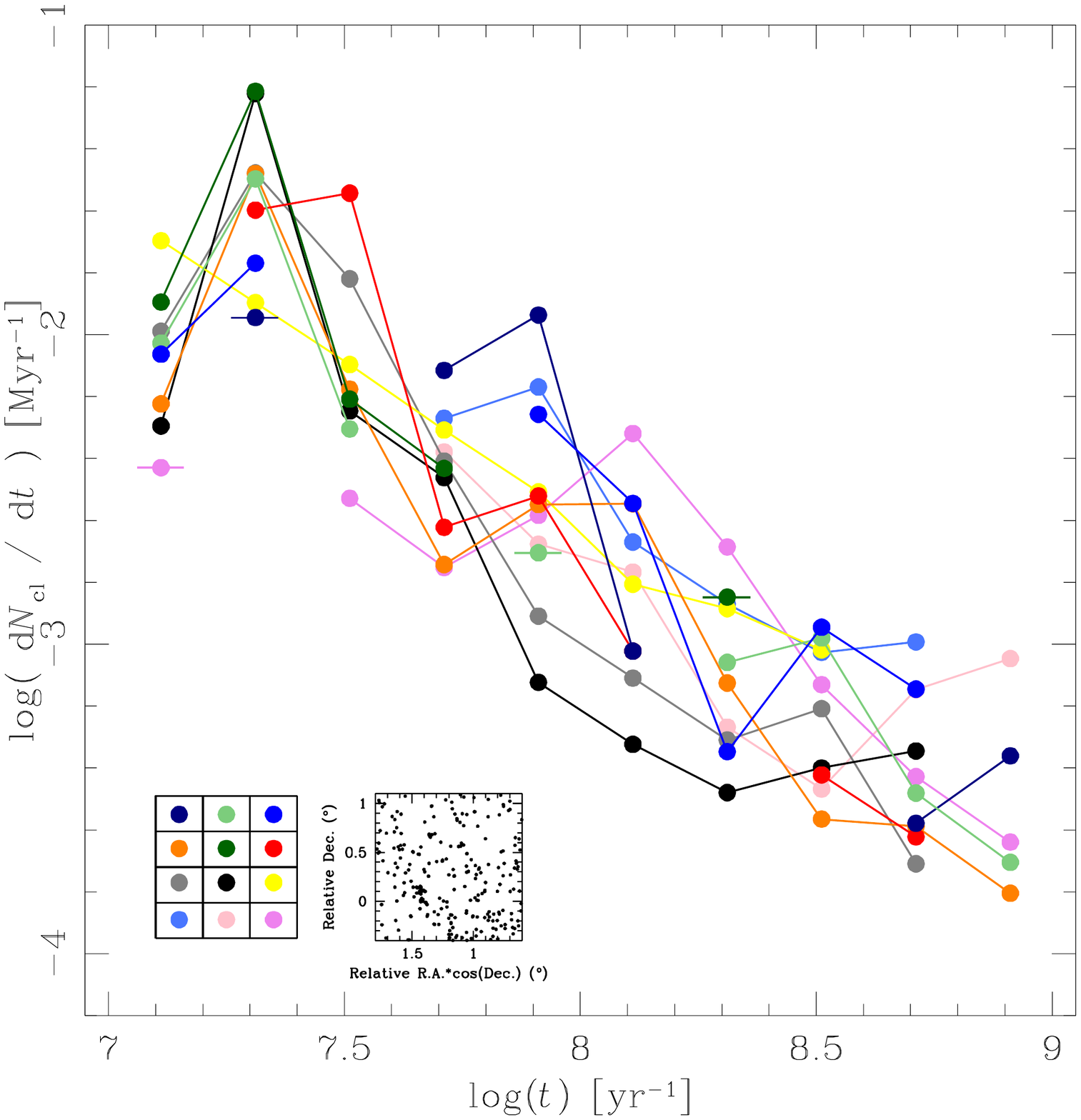}
\caption{As Fig. 6, but for tile LMC\,6\_6.}
\label{fig7}
\end{figure*}

We also compiled composite CFs using all clusters located within each
tile and compared them with those previously obtained by P14 for
HZ09's bar and the 30 Doradus region (see fig. 6 in HZ09). The CFs
were constructed by taking into account bin sizes and age
uncertainties, and normalized to the total number of clusters. Figures
8 and 9 show the resulting CFs and those from P14, represented by
black and grey lines, respectively. Both pairs of CF curves are very
similar. CF differences become important for $\log(t$ yr$^{-1})
\in[7.2,7.6]$. These differences are likely owing to the fact that
P14's CF covers the entire LMC bar, while the present CF targets its
central part only, which is characterized by a relative excess of
younger clusters with respect to the entire bar. On the other hand,
the most noticeable differences between the CFs centred on 30 Doradus
occur for older ages -- $\log(t$ yr$^{-1}) > 8.5$ -- which is most
likely also caused by the larger area covered by HZ09 than our tile
LMC\,6\_6, which thus implies that the HZ09 CF is composed of a
relatively smaller fraction of older clusters. We superimposed on
 Figs.  8 and 9 the CFs (red lines) and the corresponding uncertainties
for the bar region in HZ09 as well as their 30 Doradus region,
respectively, assuming that clusters are born according to a power-law
mass distribution with a slope $\alpha = -2$ and at a rate that is
proportional to the SFR determined by HZ09 for each region. The shapes
of the CFs recovered from the SFR curves generally follow those of the
observed CFs, although the cluster and field-star populations do not
seem to have evolved as fully coupled systems. Nevertheless, coupling
between field-star and cluster formation appears to have been dominant
for both regions. 

 Assuming that most -- if not all --
stars form in clustered modes \citep{ll03}, our results suggest
that the combination of cluster formation and disruption rates 
(which depend on environmental conditions, cluster masses, among
others) still provides a positive balance, in the sense that the
cluster population has not been entirely transformed into field stars. 
\citet{baetal13} showed that about 90$\%$ of all clusters older than 200 Myr are 
lost per dex of lifetime, which implies a cluster
dissolution rate significantly faster than that based on analytic estimates 
and N-body simulations. 
However, \citet{dgetal13} showed that there is no evidence of significant destruction, 
other than that expected from stellar dynamics and evolution in simple population
 models for ages up to 1 Gyr ($\log(t$ yr$^{-1})=9$). As can be seen,
the issue about at what pace cluster disruption takes place in different LMC regions
 and even whether
field stars mostly come from cluster disruption still needs further study.
Our results do not allow us to draw any further conclusion about this issue, but that
the cluster formation has been different in the studied regions.

By comparing our results with those of HZ09 for the two tiles, some hint about
the distint roles played by the environments can be drawn, since differences in CF are larger 
in the tile LMC\,6$\_$6 than in LMC\,6$\_$4 for ages younger than 
$\log(t$ yr$^{-1}) < 8.0$. \citet{wz11} tabulated the concentration, central
surface brightness, tidal radii, 90$\%$ enclosed luminosity radii, and local background
luminosity density for 1066 star clusters in the Magellanic Cloud Phomotometric Survey 
\citep[MCPS][]{zetal02}. They found that
while most of the star clusters are similar between the LMC and the SMC populations, the LMC lacks
star clusters that are large, either in terms of core or 90$\%$ enclosed luminosity radii as the
largest in the SMC, and suggested that this result could be featuring a signature of larger
tidal stresses in the LMC. On the other hand, \citet{wilkinsonetal03}
used N-body simulations of star clusters to investigate the possible dynamical origins of the 
observed spread in core radius among intermediate-age and old star clusters in the 
LMC and concluded that the tidal field of the LMC has not yet 
significantly influenced the evolution of the
 intermediate-age clusters. In this sense, we interpret that the actual dynamical state of 
the studied LMC clusters might be a combination of an age effect (two-body 
relataxion) and their locations in the galaxy \citep[see., e.g.][for the SMC]{getal11}.

\begin{figure}
\includegraphics[width=84mm]{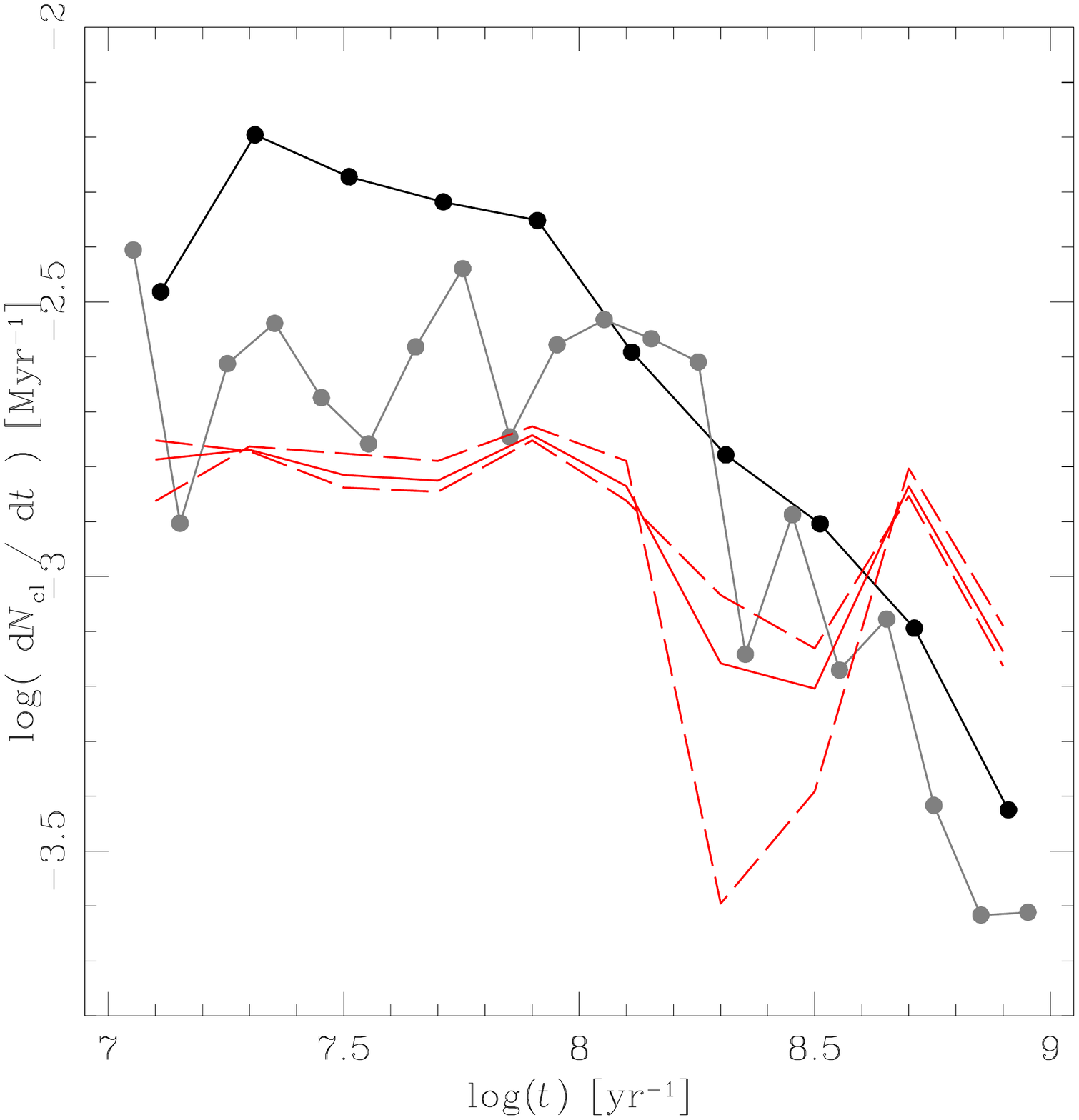}
\caption{Composite CF for tile LMC\,6\_4 (black) compared with those
  obtained by P14 for clusters within HZ09's bar region (grey) and by
  HZ09 for the respective mean SFR and its 1$\sigma$ uncertainty
  (solid and dashed red lines, respectively).}
\label{fig8}
\end{figure}

\begin{figure}
\includegraphics[width=84mm]{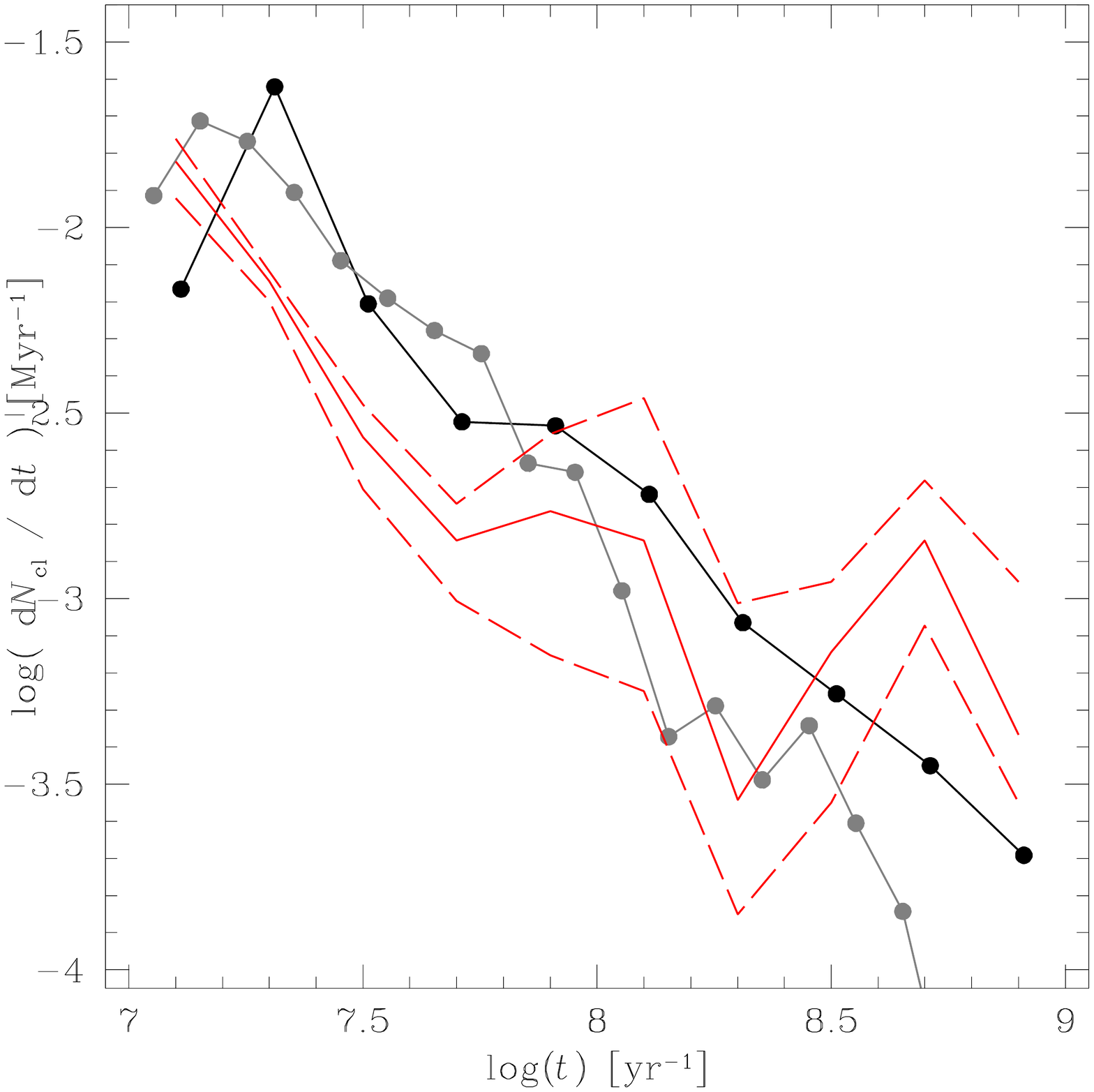}
\caption{As Fig. 8, but for tile LMC\,6\_6 and HZ09's 30 Doradus
  region.}
\label{fig9}
\end{figure}

Finally, we compared the CFs derived here with those obtained from the
SFRs derived by \citet{retal12} for tile LMC\,6\_6. The latter authors
derived the SFRs for two subregions (D1 and D2; see their fig. 2).
They used the `partial model' technique, which applies a linear
combination of partial models that optimally matches the observed Hess
diagrams allowing the distance modulus, visual extinction, age and
metallicty to vary, convolved with the distributions of photometric
errors and completeness. The coefficients of this linear combination
of partial models (including the best-fitting distance modulus and
extinction) are directly converted into the SFRs. We used their SFRs
and cluster masses from $\log(M_{\rm cl} [{\rm M}_\odot]) = 2.2$ to
$\log(M_{\rm cl} [{\rm M}_\odot]) = 5.0$ \citep{dgetal08,getal11}, and
normalized the resulting CFs by the total number of clusters used, so
that they can be compared directly with the observed
distributions. Figure 10 shows the resulting, recovered CFs, where the
uncertainties are indicated by dashed lines. The observed and
recovered CFs are clearly different for ages younger than $\log(t$
yr$^{-1}) \sim 8.0$. Even though the recovered CFs require additional
refinements, the cluster excesses could be evidence of relatively
recent enhanced cluster formation in this part of the galaxy.
Likewise, note that both recovered CFs are different between them, 
which suggests either that cluster disruption changed dramatically from one
subregion to the other, or field stars did not come from cluster disruption
in a similar proportion. \citet{retal12} and \citet{retal15}  showed that the SFR of field stars varies
with the position in the LMC and SMC, respectively. Conversely, P14 and \citet{piattietal15}
found evidence that a similar behaviour occurs within the LMC and SMC cluster populations,
but not  with a completely equivalent pace to field stars. These results, besides those
derived here, provide some clues for a better approach in the study of the 
field stars origin and its link to cluster disruption and environmental conditions.

\begin{figure}
\includegraphics[width=84mm]{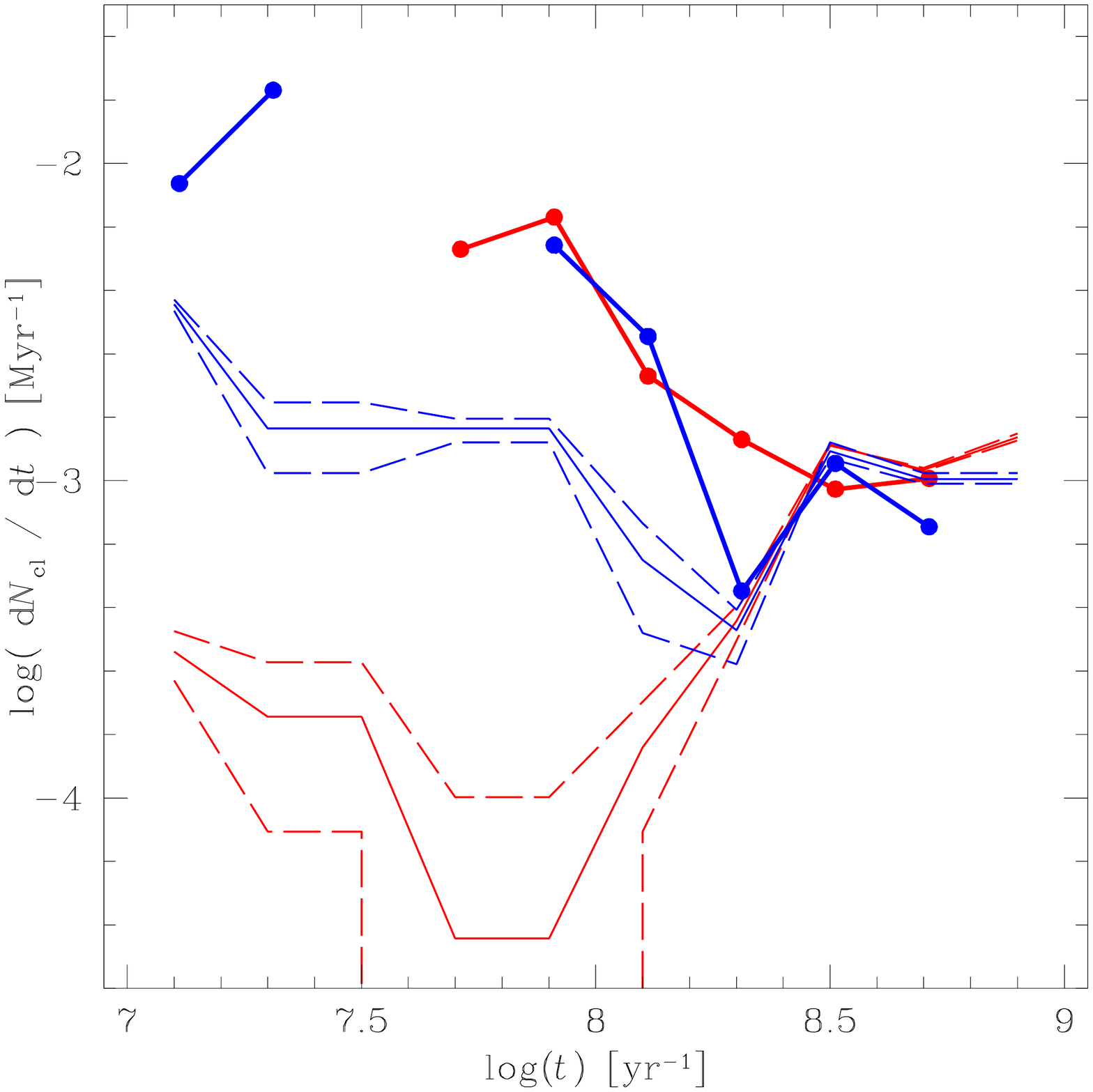}
\caption{CFs for subregions D1 (thick red line) and D2 (thick blue
  line) in tile LMC\,6\_6. Thin and dashed lines represent,
  respectively, the mean CFs and their 1$\sigma$ dispersions recovered
  from the corresponding SFRs obtained by \citet{retal12}.}
\label{fig10}
\end{figure}

\section{Conclusions}

In this paper, we have analysed CMDs of previously catalogued star
clusters located in the innermost region of the LMC based on a
$YJK_{\rm s}$ photometric data set obtained by the VMC
collaboration. We focused on tiles LMC 6$\_$4 (located in the central
part of the LMC bar) and 6\_6 (centred on 30 Doradus), because they
are among the first fully completed tiles of the VMC survey, and they
show the most recent star-formation activity in the galaxy. We
obtained PSF photometry for stars in and projected along the line of
sight towards 313 catalogued clusters, which represents a
statistically meaningful sample size.

We applied a field-star subtraction procedure to statistically clean
the cluster CMDs from field-star contamination and to disentangle
cluster features from those associated with their surrounding
fields. The technique we employed makes use of variable cells to
reproduce the field CMDs as closely as possible. Based on matching
theoretical isochrones in the VISTA system to the cleaned cluster
CMDs, we obtained estimates of the clusters' reddening values and
ages, assuming a distance modulus of 18.49 mag and a metallicity of $Z
= 0.006$ ([Fe/H] = $-0.4$ dex). Both surveyed regions are affected by
a mean reddening spanning the range $E(B-V) \in [0.2,0.3]$ mag; the
average internal (LMC) reddening amounts to $\sim$ 0.1--0.2 mag. In
addition, the tile covering 30 Doradus includes clusters with
reddening values in excess of $E(B-V)$ = 0.4 mag.

The resulting cluster ages span the age range $7.0 \le \log(t$
yr$^{-1}) < 9.0$. They form part of the cluster data base which will
result from the VMC survey and which will be used to self-consistently
study the overall cluster-formation history of the Magellanic
system. The age scale derived in this paper is consistent with those
resulting from previous publications, in particular those based on
OGLE \citep{pu00}, the Magellanic Cloud Photometric Surveys
\citep{zetal02} and the Washington photometric system
\citep{p12c,p14,chetal15}, among others. A total of 95 clusters of the
313 objects in our sample had their ages determined for the first
time. This represents an increase of $\sim$ 30 per cent of clusters
with known ages in the two LMC tile regions. The enlarged catalogue of
clusters with age estimates contains 407 objects.

Based on analysis of this statistically complete cluster sample, we
addressed the nature of variations in the LMC CF in terms of the
clusters' spatial distribution. Particular attention was paid to bin
sizes and age uncertainties in constructing CFs for 12 different
subregions of similar sizes, homogeneously distributed within each LMC
tile. As for tile LMC\,6\_4, we found that the outermost subregions of
the LMC bar experienced enhanced cluster formation activity in the
past -- $\log(t$ yr$^{-1}) \in [8.5,9.0]$ -- whereas more recently,
$\log(t$ yr$^{-1}) < 7.7$, cluster formation seems to occur more
frequently in the inner regions. Overall, however, cluster formation
in the central parts of the bar has proceeded in a patchy manner. On
the other hand, in tile LMC\,6\_6, the 30 Doradus subregions play the
most striking role in the very recent cluster formation history. They
are surrounded by  successively older clusters for increasing
  distances from 30 Doradus.

Finally, we compared the composite CFs for each LMC tile to those
recovered assuming that clusters have formation rates similar to the
known field SFRs. The shapes of the recovered CFs, and more
particularly those for the subregions near 30 Doradus, suggest that
cluster and field-star populations do not seem to have evolved
as fully coupled systems during the last $\sim$ 100 Myr. The cluster
excesses found could be evidence of relatively recent enhanced cluster
formation in these parts of the galaxy and thus confirm both that most
stars are formed in clusters and that there is no significant
destruction other than that expected from stellar dynamics and
evolution of simple stellar population models for ages up to 1 Gyr.

\section*{Acknowledgements}
We thank the Cambridge Astronomy Survey Unit (CASU) and the Wide-Field
Astronomy Unit (WFAU) in Edinburgh for providing calibrated data
products under the support of the Science and Technology Facilities
Council (STFC) in the UK. This research has made use of the SIMBAD
data base, operated at CDS, Strasbourg, France.  We thank the
  anonymous referee whose comments and suggestions helped us to
  improve the manuscript. RdG acknowledges research support from the
National Natural Science Foundation of China (NSFC; grant 11373010).

%to be commented before sending to editor
%\bibliographystyle{mn2e_new} %style mn.bst
%\bibliography{paper} % your references file.bib
% 
%to be uncommented before sending to editor

%

%\clearpage

%\setcounter{figure}{1}
%\begin{figure*}
%\fbox{\includegraphics[width=50mm]{cmd_yk_smc35.pdf}}
%\fbox{\includegraphics[width=50mm]{cmd_yk_smc45.pdf}}
%\fbox{\includegraphics[width=50mm]{cmd_yk_smc65.pdf}}\\
%\fbox{\includegraphics[width=50mm]{cmd_yk_smc56.pdf}}
%\fbox{\includegraphics[width=50mm]{cmd_yk_bri23.pdf}}\\
%\caption{CMDs for stars in tiles SMC 3$\_$5, 4$\_$5, 6$\_$5, 5$\_$6 and BRI 2$\_$3 with error bars
%coloured according to the colour scale of the completeness level.}
%\label{fig2}
%\end{figure*}

\label{lastpage}
\end{document}